\documentstyle[emulateapj]{article}
\def\ie{i.e.\ }
\def\eg{e.g.,\ }
\def\etal{et~al.\ }
\def\ltsima{$\; \buildrel < \over \sim \;$}
\def\simlt{\lower.5ex\hbox{\ltsima}}
\def\gtsima{$\; \buildrel > \over \sim \;$}
\def\simgt{\lower.5ex\hbox{\gtsima}}

\def\h1{\ion{H}{1}\ }

\def\h2{H$_2$}
\def\coh2{CO/H$_2$}
\journalid{}{}
\articleid{}{}
\lefthead{Mihos \etal}
\righthead{Models of Cuspy Triaxial Galaxies}
\slugcomment{Submitted to the {\it Astrophysical Journal}}
 
\begin{document}
 
\title{Models of Cuspy Triaxial Galaxies}
 
\author{Kelly Holley-Bockelmann,\altaffilmark{1} J. Christopher Mihos,\altaffilmark{1}
Steinn Sigurdsson,\altaffilmark{2} and Lars Hernquist\altaffilmark{3}}

\altaffiltext{1}{Department of Astronomy, Case Western Reserve University,
10900 Euclid Ave, Cleveland, OH 44106, hos@burro.astr.cwru.edu,
kelly@eor.astr.cwru.edu}
\altaffiltext{2}{Department of Astronomy and Astrophysics,
Penn State University, 525 Davey Lab,
University Park, PA 16802, steinn@astro.psu.edu}
\altaffiltext{3}{Department of Astronomy, Harvard University,
60 Garden St, Cambridge, MA 02138, lhernqui@cfa.harvard.edu}
 
\begin{abstract}

We construct numerical models of mildly triaxial elliptical galaxies with 
central density cusps. Using a technique we call ``adiabatic squeezing,''
we begin with a spherical $\gamma=1$ Hernquist model and apply a drag to
the velocities of the particles along each principle axis. The final models 
are stable in isolation, preserving their density structure and figure 
shape over many dynamical timescales. The density profile and axial
ratios compare well to the observed properties of elliptical galaxies.
The orbital structure of these models show a mixture of tubes, boxes,
and boxlets, as expected for triaxial systems, with very few chaotic
orbits. These $N$-body realizations of cuspy triaxial galaxies provide a 
basis for the study of dynamical evolution of elliptical galaxies.

\end{abstract}
 
 
\section{Introduction}

There is good reason to believe that many elliptical galaxies are at least
moderately triaxial systems. Observationally, the distribution of
isophotal shapes of ellipticals argues that there are few perfectly
spherical or axisymmetric members of the present-day elliptical
population (Tremblay \& Merritt 1995; Ryden 1996). Using morphological
and kinematic data, Franx, Illingworth, \& de Zeeuw (1991) have shown
that ellipticals show mild triaxiality, while more recently Bak \&
Statler (2000) have arrived at a similar conclusion by modeling the
kinematics of ellipticals.  There is strong theoretical support for
triaxiality as well. Under the gravitational collapse model for
elliptical galaxy formation, models of protogalaxies undergo a
non-spherical collapse which results in a virialized system that is
decidedly triaxial (Dubinski \& Carlberg, 1991). 
Similarly, in the hierarchical merging picture,
the remnants of simulated galaxy mergers also show strong triaxiality
(\eg Hernquist 1992; Barnes 1988, 1992; Hernquist 1993), even when multiple mergers are considered (Weil \& Hernquist 1996). Either model of elliptical 
galaxy formation
predicts that a dynamically young elliptical should possess a fair
degree of triaxiality, although subsequent evolution may drive the
system towards axisymmetry.

There is also good reason to believe that many ellipticals have cuspy
density profiles. While elliptical galaxies were once characterized by 
flat, isothermal cores, studies using the {\it Hubble Space
Telescope} have shown that in fact the density profile of ellipticals
rarely becomes perfectly flat in the inner region. Instead, the density
profile continues to rise as $\rho \propto r^{-\gamma}$, where $0.25 <
\gamma < 2$ (\eg Lauer \etal 1995; Byun \etal 1996; Gebhardt \etal
1996; Faber \etal 1997).  Furthermore, correlations exist between the
cusp slope $\gamma$ and the global properties of the galaxy, including
shape in the form of boxy or disky isophotes (Faber \etal
1997). Dynamical arguments suggest also that the presence of a strong
central cusp ($\gamma > 1$) acts to drive chaos in the orbit families
which populate the galaxy, driving the system away from strong
triaxiality (\eg Gerhard \& Binney 1985; Norman, May, \& van Albada
1985; Merritt \& Valluri 1996; Holley-Bockelmann \etal 2000).

While real elliptical galaxies most probably are cuspy and moderately
triaxial, most self-consistent modeling of ellipticals to date has
focussed either on cuspy spherical or axisymmetric models, or triaxial
models with constant density cores (\ie $\gamma=0$). The perfect
ellipsoid of de Zeeuw (1985) allows for differing degrees of
triaxiality, but has a flat $\gamma = 0$ core.  Self-consistent $N$-body
models of elliptical galaxies used in studies of black hole growth
have either examined triaxiality in a $\gamma \sim 0$ profile (\eg Norman
\etal 1985; Merritt \& Quinlan 1998), or focused on cuspy models which
are spherical (Sigurdsson \etal 1995; Quinlan \& Hernquist, 1997) or 
axisymmetric (van der Marel \etal 1997). The problem lies in the lack 
of an analytic distribution function for systems which are both cuspy 
and triaxial; without such a distribution function, generating initial 
conditions for simulations is difficult.

Here we employ an alternative approach to generating triaxial galaxy
models with well-defined shapes and cusps. We apply an adiabatic drag
on the particles in a spherical $\gamma=1$ Hernquist model to mold the
system into triaxial shape. This process preserves the cusp slope and
results in models which are moderately triaxial and have axis ratios
that are reasonably constant with radius. More importantly, the models
are stable, exhibiting little evolution in their structural properties
over many crossing times. We characterize the orbit families which
populate such models, and find a rich mixture of tubes, boxes, and
boxlets, with very few chaotic orbits. This technique has
also been applied successfully in a $\gamma = 0$ model, and
can, in principle, generate models with a wide range in cusp slopes.
These well-defined models
thus present a useful tool for studying the evolution of
elliptical galaxies or triaxial halos.

\section{Modeling Technique}

Because there is no known distribution function for triaxial
elliptical galaxies, we cannot simply sample an analytic distribution
function to generate initial conditions for the model. Instead, we
begin with an object with a known distribution function -- a
spherical, isotropic model which has a distribution function $f(E)$
which is a function only of energy -- and mold that model into a
triaxial shape by applying an artificial drag on the particles.

We begin with a spherical Hernquist (1990) model, which has the
density profile
$$\rho(r)={M\over 2\pi}{a\over r}{1\over (r+a)^3}$$ where $M$ is the
total mass and $a$ is a scale length. In projection, this model
follows closely an $R^{1\over 4}$ law, and possesses a
three-dimensional central density cusp of the form $\rho \sim r^{-1}$
as $r \rightarrow$ 0.  This model belongs to a family of $\gamma$ models
(Dehnen, 1993; Tremaine \etal 1994) whose density profile is given by
$$\rho_{\gamma}(r)= {{(3-\gamma)M}\over{4\pi a^3}} {\left({{r}\over{a}}\right)^{-\gamma}} {{\left(1+ {{r}\over{a}}\right)^{-(4-\gamma)}}} ,$$ such that the
Hernquist profile corresponds to $\gamma=1$.  The Hernquist model
has the distribution function (Hernquist 1990)
\begin{eqnarray*}
  f(E) & = & {M\over 8\sqrt{2}\pi^3a^3v_g^3}{1\over (1-q^2)^{5/2}} [3
       \sin^{-1} q + \\ & & q(1-q^2)^{1/2}(1-2q^2)(8q^4-8q^2-3)]
\end{eqnarray*}
where
$$q=\sqrt{-{a\over GM}E}$$ and
$$v_g = \left({GM\over a}\right)^{1/2}.$$ To generate the initial
spherical model we sample this distribution using a multimass scheme
developed by Sigurdsson \etal (1995), wherein particles have a mass
which is roughly inversely proportional to their pericentric radius. In this
manner, in the central regions of the model the particle number
density rises faster than the mass density, permitting better
resolution and sampling of the phase space distribution of the nuclear
region. Because of the SCF method of force calculation (see below),
this multi-mass technique does not introduce any spurious mass
segregation or relaxation into the model.

One the initial model is constructed, we apply a time-varying
adiabatic drag on the motions of the particles to mold the system into
a triaxial shape. The construction is a three step process. A drag is
first applied to the $z$ axis of the model while forcing axisymmetry
on the system through the zeroing of odd terms in the SCF expansion of
the gravitational potential. This drag term is smoothly turned on as
$$\xi(t) = \xi_0 \left[ 3\left({t\over t_{grow}}\right)^2 -
2\left({t\over t_{grow}}\right)^3 \right]$$ where $\xi$ is the drag
factor applied to the velocities and $t_{grow}$ is the timescale over
which the drag grows. This form results in a smooth development of the
drag factor, as $\dot{\xi}=0$ at $t=0$ and $t=t_{grow}$. The drag is
then used to modify particle velocities as
$$\vec{v}' = \vec{v}\left[{(1-{1\over 2}\xi\Delta t)\over(1+{1\over
2}\xi\Delta t)}\right]$$ which ensures a smooth onset and termination
of the drag coefficients.  After $t=t_{grow}$, the drag remains at
full strength for a time $t_{drag}$, after which it is slowly turned
off over a time $t_{decay}$.  For the models discussed here, we
employed $t_{grow}=t_{decay}=10$ and $t_{drag}=30$, compared to a
half-mass dynamical timescale for the model of $t_{dyn, 1/2} = 8.33$.

During the squeezing process, the model contracts in radius as the
system adjusts to the induced change in binding energy. Once the z-dragging is
complete, the radius and velocity vectors of the particles are
rescaled so that the system is in equilibrium with scale radius $a=1$.
The axisymmetry requirement is then relaxed and dragging ensues along
the y-axis on similar timescales, but with a different drag
coefficient ($\xi_0$). When y-dragging is complete, the system is
again rescaled to $a=1$ and is evolved forward in time with all
dragging shut off, allowing it to settle into an equilibrium
configuration.

The models are evolved using the self-consistent field (SCF) method of
Hernquist \& Ostriker (1992). The SCF method expresses the density and
potential as an expansion in a set of basis functions, the lowest
order of which represents the spherical Hernquist density profile. The
expansion coefficients are determined from the particle distribution,
using $n$ radial terms and $l,m$ angular terms. For the high
resolution $N=512,000$ particle models shown in the next section we
use $n_{max}=10, m_{max}=l_{max}=6$, while for smaller models
($N=128,000$) we use $n_{max}=6, m_{max}=l_{max}=4$ to reduce root-N
noise in the determination of the coefficients. The orbits of the
particles are integrated using a high-order hermite integrator with
variable time steps.

\section{Generating a Triaxial Model}

We demonstrate this process with a N=512,000 particle $M=a=1$
Hernquist model.  For this model, we use squeezing coefficients
$\xi_{0,z} = 0.03$ and $\xi_{0,y} = 0.025$, with
$t_{grow}=t_{decay}=10 {\ \rm and\ }t_{drag}=30$.

During the squeezing, the system retains its initial density profile
(once the model is rescaled so that $a=1$). Figure 1 shows the density
profile of the model throughout the squeezing process. The profile is
virtually unchanged from its original shape, save for a minor ``kink''
in the density profile at large radius ($r\sim 50$). Certainly out at
this distance the dynamical timescale is significantly longer than the
dragging timescale, and these particles simply have not responded as
strongly to squeezing. The density is extremely low at this distance,
however; only 4\% of the total mass is found outside this radius. The
bulk of the system is well withing this radius, where the density
profile is very well behaved.  To amplify this, we show in Figure 2
the slope of the central density cusp as a function of time. Again
there is very little evolution away from the original $\gamma=1$
slope. Evaluated at $\log r = -2$, the slope does show a gradual
flattening over the course of the simulation from $\gamma=1$ to
$\gamma=0.9$; however this evolution is consistent with the effects of
wandering of the expansion centroid due to root-N fluctuations in the
expansion. The low N model shows stronger
evolution, as would be expected if the change in cusp slope was driven
by particle noise. This small evolution aside, the model exhibits an
unchanging density profile over $\sim 5$ orders of magnitude during
and after the squeezing process.

The evolution in figure shape is shown in Figure 3, evaluated at three
different ellipsoidal radii. As the first phase of squeezing begins
along the $z$ axis, the model becomes very oblate ($c/a \sim
0.6$). Once squeezing begins along the intermediate axis, the model
becomes triaxial, although the short axis rebounds a bit so that when
squeezing is complete the final model is only moderately triaxial,
with axis ratios 1:0.85:0.7. The rebound of the minor axis may reflect
the incompatibility of a $\gamma=1$ cusp slope with a strongly
triaxial system (\eg Merritt 1997; Valluri \& Merritt 1998), as the box
orbits which support strong triaxiality become chaotic in a the
presence of a strong central cusp. It is not clear that this
explanation is correct; given the non-physical drag and rapidly
changing potential, other effects may be in play. To some degree,
however, understanding the detailed dynamics at these intermediate
stages is rather ill-motivated given the non-physical adiabatic drag
we apply. Of more importance is the post-drag equilibrium
configuration, which we explore in more depth in the next section.

To explore the variety of shapes which could be constructed using this
method, a suite of smaller ($N=128,000$) models was also calculated in
which the dragging coefficients were varied. Table 1 gives the drag
coefficients and final shape of these models, while Figure 4 shows the
models on the triaxiality plot. The models populate a region of
moderate triaxiality, with shapes which range from somewhat oblate to
somewhat prolate systems.  Ideally, there would be a unique mapping of
drag coefficients to final figure shape; this unfortunately is not the
case. As such it remains difficult to predict the final shape for a
given set of drag coefficients. Nonetheless, the calculations do
present a set of stable, triaxial, and self-consistent $\gamma=1$
galaxy models useful for studying the evolution of triaxial
ellipticals.

\section{Internal Structure}

Here we look at the internal structure of the high resolution fiducial
model in more detail. Figure 5 shows the properties of the model as a
function of ellipsoidal radius $q$ ($=\sqrt{x^2 + (y/b)^2 + (z/c)^2}$)
at a time $t=150$, or $7.5t_{dyn}$ after squeezing has terminated. For
the plots of density and velocity dispersion (Figures 5a,c,d), $q$ is
calculated using a constant $(b,c)$ evaluated at the half mass
radius. To calculate shape as a function of radius (Figure 5b), we
use a technique described by Dubinski \& Carlberg (1991), wherein
particles are first binned into a spherical shell at a given radius,
and the moment of inertia tensor for the binned particle distribution
is used to estimate the ellipticity of the particles. The bin shape is
then adjusted based on the estimated ellipticity, and the procedure
continues iteratively until the bin shape and the ellipticity match.

Figure 5a shows that the model retains its $\gamma=1$
character over many orders of magnitude in radius. The model also has
fairly constant shape (Figure 5b), although there is a slight trend
for the outer regions of the model to be more axisymmetric than the
inner regions. This probably reflects the fact that the squeezing
timescale was independent of radius, so that the squeezing lasted for
a shorter dynamical timescale at large radius.  In the outskirts of
the model, the particles have not responded as strongly as those in
the inner regions.

The projected velocity dispersion (Figure 5c) shows a rollover at
small radius, as is to be expected from the Hernquist distribution
function (Hernquist 1990). The velocity ellipsoid reflects the
triaxial shape of the model, with $\sigma_x > \sigma_y > \sigma_z$,
where $x$ and $z$ are the major and minor axes of the model,
respectively. The model also displays a
slight radial anisotropy in the velocity ellipsoid (Figure 5d; $\beta
= 1-\langle v_t^2\rangle / \langle v_r^2\rangle > 0$). 
At large radius ($\log q > 1$) the model becomes
strongly anisotropic; for these particles the squeezing has not
been adiabatic, as they experienced the squeezing for a fraction of an
orbital timescale. In the outer regions of the model, the system has
not yet come into full equilibrium, and the model is somewhat suspect.
However, the energy and angular momentum content of these particles is
large, and they will not greatly affect the equilibrium dynamics in
the inner regions of the model. Indeed, the stability of the density
profile and shape support this conjecture.

\section{Orbital Properties}
Since the intrinsic shape of any galaxy is dictated by the time
average of its orbital content, orbital structure analysis is an
essential probe of the viability and stability of any galaxy
model. For example, triaxiality in galaxies is supported by the large
fraction of stars on box orbits (or boxlets) as determined by the
shape and density profile of the potential (e.g., Schwarzschild 1979).
The fact that box orbits can travel arbitrarily close to the potential
center argues that steep density cusps may destabilize these orbits,
seeding chaotic motion and driving galaxies away from triaxiality 
(Merritt 1997). Indeed, Merritt finds
that $\gamma=1$ represents a critical profile, and suggest that
galaxies with steeper cusps should quickly evolve away from
triaxiality.  Given that our models represent moderately triaxial
systems with $\gamma=1$ cusps, it is interesting to characterize their
allowed orbit families and level of chaos. In fact, as we demonstrate
below, the equilibrium state of these models displays a mixture of
tubes, boxes, and resonant orbits, and are not strongly chaotic.

Characterizing the orbital content of these models also allows us to
lay the groundwork for future studies of evolution in triaxial
ellipticals.  The growth of a massive black 
hole in a triaxial potential can destabilize 
centrophilic box orbits through stochastic diffusion, driving 
the global shape of a galaxy toward axisymmetry in a few 
crossing times (Gerhard \& Binney 1985; Norman, May, \& van Albada 1985;
Merritt \& Quinlan 1998; Wachlin \& Ferraz-Mello 1998; 
Merritt \& Valluri 1998). The
change induced in the orbital content (and hence the galaxy structure)
by a central black hole will thus be very sensitive to the initial
orbit content of the triaxial system. As a case in point, strong shape
evolution could be tempered by a population of centrophobic boxlets
which could avoid scattering by the black hole, thereby maintaining
triaxiality.  An analysis of the orbit populations in these triaxial
models will help address this possibility, and provide an important
contrast to the orbital families present under the influence of a
central black hole (Holley-Bockelmann etal 2001). 
For this exercise, it is necessary to determine the initial
orbital content and set up the technique here.

Orbit analysis has traditionally been explored in a static, analytic 
potential.  Since
an N-body representation of the final potential can scatter orbits
artificially, it is crucial to reduce the coarse-grained nature of the
final model as much as possible. Fortunately, a triaxial figure has a
mirror symmetry about each axis, such that a particle at {$x, y, z, v_x,
v_y, v_z$} could also exist at {$-x, y, z, -v_x, v_y, v_z$} without distorting
the potential (and so on for each axis). We take advantage of this
8-fold symmetry by seeding the final state of the model with these
symmetric pseudo particles. The SCF
code was then used with these pseudo particles to obtain the coefficients for the potential
expansion.  With the effective particle number increased from 512,000
to over $4 \times 10^6$, the noise in the potential was drastically
reduced.  Hence, all the orbits in this discussion were evolved in the
frozen potential dictated by the expansion coefficients from this
symmetrically seeded final state.

\subsection{Orbit Classification Techniques}

The large number of particles in each model demands an automated
classification technique. In this subsection, we will describe 
our classification scheme first for 2-d and then for 3-d orbits.
In general, we designed a method
based on two common classification techniques:
axis-crossing pattern recognition and fourier spectral analysis.

The axis-crossing technique relies on the fact that a regular orbit
displays resonances that are recognizable by the pattern which emerges
as the particle crosses an axis along its orbit (Fulton \& Barnes
2000).  For example, in the simplest case, a tube orbit confined to
the $xy$ plane crosses first the $x$-axis, then the $y$-axis, and will
repeat exactly that pattern ad infinitum. Therefore, its axis-crossing
pattern will be: $xyxy$.... A boxlet commonly called a ``fish'' will
have the pattern $xyxyxxyxyx$...; in fact every planar resonant orbit
will have a characteristic axis-crossing pattern. Planar box orbits, on the
other hand, are characterized by their lack of a regular axis-crossing
pattern. This technique is excellent for the analysis of $N$-body generated
orbits, because the act of crossing an axis is insensitive to noise in
the potential. Unfortunately, this technique is not useful in
discriminating between a pure box orbit and a chaotic orbit, since
neither orbit has a unique axis-crossing pattern.

Classification by fourier spectral analysis is similar in spirit to
the axis-crossing technique in that it looks for the characteristic
patterns in fourier space that are generated by a regular orbit.  For
a regular orbit, the fourier transform of the time sequence for each
degree of freedom results in a spectrum with frequencies that
correspond to the quasi-periodic motion of the particle, and the
frequencies with the maximum amplitude for each coordinate are the
dominant frequencies. 

In two-dimensions, our technique was based loosely on the fourier
classification scheme of Carpintero \& Aguilar (1998). In short, from
the fourier transform of each coordinate, the orbit is classified by
resonances between each pair of coordinates, such that $ \mid m f_1 -
n f_2 \mid > \epsilon $, where m and n are integers, $f_1$ and $f_2$
are the dominant frequencies, and $\epsilon$ is an accuracy parameter
which depends on the frequency resolution and orbital accuracy (due to
potential noise and integration errors).  For example, the $xy$ planar
tube orbit will exhibit a 1:1 resonance between the dominant
frequencies in the $x$ and $y$ direction, while fish boxlets will have a
3:2 resonance. Planar box orbits will contain many peak frequencies that are
nearly as large as the dominant frequency, so box orbits can be
identified both by their resonance and by the number of significant
peaks.

The classification of three-dimensional orbits is not a trivial
'scaling-up' of these 2-dimensional methods. Since many 3-d 
resonant orbits do not project to an identifiable planar 
periodic orbit in any plane,
the axis-crossing technique may misidentify a resonant orbit (i.e.
a (2:1:-2) orbit) as a box (Merritt \& Valluri 1999). This 
misidentification is important because resonant orbits 
avoid the center, and are thus thought to be more stable against
a central density cusp or black hole than a centrophilic box.
Fortunately, any stable resonant orbit still obeys a fixed ratio between 
the dominant frequencies in each coordinate, so on a frequency map
(a plot of $f_x / f_z$ vs. $ f_y / f_z$) resonant
orbits stand out as a complex grid - the slope and intercept 
of any line on this grid identifies the particular resonance 
(Laskar 1993, Merritt \& Valluri 1998).

Aside from the ability to identify non-planar 3-d resonant orbits, the 
true advantage of the fourier technique is that it can easily
quantify the stability of any orbit. Since the dominant frequency in a
chaotic orbit will change with time, it is simple, in principle, to
determine the number of chaotic orbits in a given potential.
Stochasticity is calculated by the change in the dominant frequency
over successive time intervals. So, while the axis-crossing technique
cannot differentiate between boxes and chaotic orbits, this fourier
method easily identifies the chaotic orbit as one with a changing
dominant frequency. Unfortunately, the fourier technique is sensitive
to noise in a potential, so it is unable to detect a subtle onset of
chaos. However, since we are interested in the global stability of our
models over many dynamical times, it is sufficient to detect only the
stochasticity that is strong enough to change the bulk properties of
the model over this time.

\subsection{Methods and Results}

There are two different questions that can be addressed: (1) what
orbits are possible in a given potential, and (2) what orbits are
actually populated by the particles. The first question can be
explored with an analysis of the surface of section, a 2-d technique,
while the second question must focus on a fully 3-d analysis.  We
addressed both questions, and this section describes the technique and
results of each.

\subsubsection{2-D Surfaces of Section}

We integrated orbits with initial conditions designed to evenly trace
the phase space along the $xy$ and $xz$ planes. There are four sets of
$O(10^4)$ particles for each plane, corresponding to binding energies
of $E=-1.0, -0.65,-0.40, -0.20$ and typical mean radii of $\langle q \rangle = 0.2, 
0.6, 1.3, 3.9$. The particles are followed in the
frozen final potential, and are confined to the initial plane to
reduce scattering by noise.

To classify the 2-d orbits confined to the $xy$ and $xz$ planes, we
integrated the particles for up to a total of ~200 dynamical times, or 
until T=4000, whichever occurred first. In sets
of ~50 dynamical times, the orbit was equally sampled 524288 times and
the fourier transform of this time series was calculated by a FFT with
a Hann window function. The first two fourier series were used to
determine two independent orbital classifications, and each subsequent
time series was used only to determine if the orbit had become
chaotic, where a strongly chaotic orbit was defined in the convention of
Valluri \& Merritt (1998) as: $\Delta f > 0.1$.  We define $\Delta f =
\mid f_1 - f_2 \mid/ f_0 T$, where $f_1$ and $f_2$ are the dominant
frequencies at the first and second time intervals, $f_0$ is the
frequency of a tube about the long axis, and T is the time interval.
That is not to say that orbits with $0 < \Delta f < 0.1$ are not
chaotic; however, the diffusion timescales of these orbits were too
slow for noticeable changes in the shape over 200 dynamical times.
Since we are interested only in stochasticity strong enough to induce
bulk evolution of the model, we can neglect more subtly chaotic
orbits (see also Valluri \& Merritt (1998) for a discussion).
The coarse-grained potential also prevented us from distinguishing
subtly chaotic orbits from slightly noisy ones, so the detection of
mildly chaotic orbits is not reliable.

To complement the fourier classification, each orbit was also
classified according to its axis crossing pattern. Up to 100 axis
crossings were counted to compile the axis-crossing pattern, and were
compared to template patterns with resonances up to 7:6. The higher
resonances were relatively insignificant in this model, so they were
not tracked. Approximately 200 orbits were cross-checked by visual
identification to determine the accuracy of the classification method
and to tune the technique, and less than $ 1 \% $ of the orbits were
misidentified in each subsample.

Figure 6 shows the surface of section for each energy slice along the
$xy$ and $xz$ planes. Notice the profusion of boxes and lower-order
boxlets in the most bound sets of orbits, characteristic of a cuspy
triaxial potential (\eg Miralda-Escude \& Schwarzschild 1989). In our
model, boxlets comprise a larger fraction of phase space along the
$xz$ plane than along the $xy$ plane. Among boxlets, fish orbits (3:2)
are the dominant boxlet on the $xz$ plane, and pretzels (4:3) are
dominant on the $xy$ plane. A common boxlet in many analytic
potentials, bananas (2:1), are an insignificant part of this model,
most probably due to the relatively modest flattening of the figure
along each axis (Fridman \& Merritt 1997). On either plane, it is 
clear that tubes occupy a
larger amount of phase space in the outer parts of the potential.
In addition, less than $0.2 \%$ of planar orbits were stochastic,
as defined above.

\subsubsection{Three-Dimensional Population}

It is also important to know the actual mixture of orbits traced by
the particles in the $N$-body model. With this in mind, we initialized
the orbits directly from the final state of the $N$-body model. The
pool of 512,000 particles were sorted according
to binding energy and binned in 9 energy slices. As before, particles
in the slices corresponding to binding energies $E=-1.0, -0.65,-0.40,
-0.20$ were followed in the final frozen potential, 
although the orbits were allowed to move in all three dimensions. 
Since the particles are not confined to a plane, the three 
dimensional orbits were much more
susceptible to noise in the coarse-grained potential, which makes
accurate classification more difficult. 

Nonetheless, using the same fourier routine, integration length,
and sampling parameters, we were able to extract the fundamental frequencies
from these orbits, and identify the major resonances using 
the 3-d technique outlined above. Figure 7 
presents the frequency map for each energy slice. Notice the strong
(1:-2:1) resonance, which persists from the smallest to the largest radii.
In the first two energy slices, many planar boxlets are visible, 
both in the region around $(f_x/f_z,f_y/f_z)=(0.8,1.0)$, and in the clump 
around (0.7,0.9). The region at $\approx$ (0.8,1.0) corresponds to 
orbits that project to tubes in one direction and low order boxlets, 
like fish, in another. The clump at (0.7,0.9) contains a confluence 
of planar boxlets, low-order 3-d resonances, boxes (which are non-resonant and
therefore fill regions off a grid of resonant lines), and orbits
that could be identified as either mildly chaotic or noise dominated.
In the outer slices, most planar boxlets all but disappear, though the 
resonant orbit (2,0,-3) is still present. As in the surface of section, the 
gradual take-over by tubes is apparent in the outer parts of the model, 
as we go from nearly $60 \%$ tubes (by mass) in the inner region to over 
$90\%$ in the outermost slice. 

Since a frequency map represents the frequency ratios at
a single time, it is not possible to accurately identify stochastic orbits
using this plot. Stochasticity for these 3-d orbits 
was determined in the same manner as the planar orbits, by determining
the change in the dominant frequency over the integration interval. 
In this stable 
model, we observe less than $ 1 \% $ of the particles on substantially
stochastic orbits. In fact, the large fraction of tubes vs. 
boxes in our models, a reflection of the modest triaxiality, may preclude the 
existence of many chaotic orbits. Indeed, nearly every 
stochastic orbit was found in the innermost
slice, where noise from the potential dominates; thus it is 
not clear that even these apparently stochastic orbits represent 
the chaotic orbits found in strongly cuspy, strongly triaxial 
figures (\eg Merritt 1997), where boxes are a mainstay. The
small number of stochastic orbits found in our model may simply have 
resulted from noise in the potential.

\section{Summary}

Using a technique we refer to as ``adiabatic squeezing'' we have constructed
models of triaxial galaxies with central density cusps. These models
preserve a Hernquist (1990) $\gamma=1$ profile over many orders of magnitude,
from $-2 < \log(r/a) < 2$, or 10 pc $ < r < $ 100 kpc when scaled to
a luminous elliptical with half mass radius 1.5 kpc. The models possess
fairly constant axis ratios as a function of ellipsoidal radius, showing
the range of triaxiality ($0.3 < T < 0.75$) and modest flattening inferred
for massive ellipticals.
The models are stable in isolation, showing little evolution in either the
density profile or figure shape over many dynamical times.

A combination of fourier spectral classification and axis-crossing
pattern recognition have been used to classify the orbit content of a
characteristic model. At the most tightly bound energies, the models
possess a rich variety of orbits, including tubes, boxes, and resonant
boxlets, reflecting the triaxial nature and central cusp of the
model. Moving to progressively less bound energy slices, the ratio of
loops:boxes rises such that the outer regions contain predominantly
loop orbits. While the presence of noise in the potential expansion
limits our ability to detect modest chaos in the orbit populations, we
find no strongly chaotic orbits in the model of the type that would
result in rapid evolution of the model.

These models thus represent a useful tool for studying the
self-consistent dynamical evolution of triaxial galaxies and halos. We
are currently using these techniques to explore the evolution of
elliptical galaxies harboring central massive black holes
(Holley-Bockelmann \etal 2001). Other potential uses include studying
the inflow of gas in triaxial systems, the response of triaxial dark
halos to baryonic accretion and disk formation, and the infall of
satellite companions into elliptical galaxies. 

\acknowledgements

This work is supported through a grant of computing time from the
National Partnership for Advanced Computational Infrastructure and the
San Diego Supercomputer Center, and by NASA through grants NAG5-7019
and HF-01074.01-94A. We thank Colin Norman for many useful
discussions, and Rebecca Stanek for help with data analysis.
 

\clearpage

\begin{figure}
\plotone{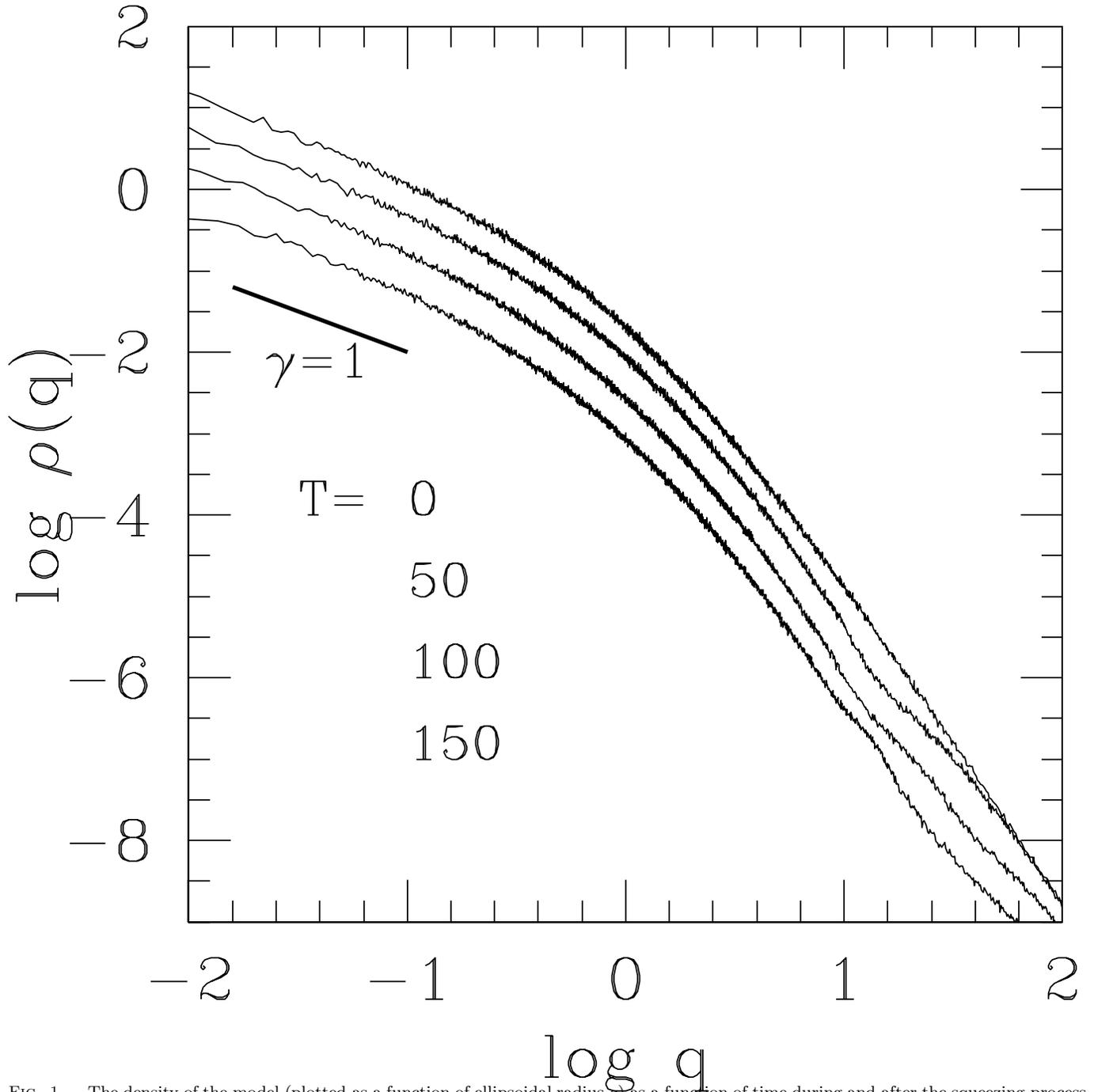}
\caption{The density of the model (plotted as a function of ellipsoidal
radius $q$) as a function of time during and after the squeezing process.
The curves show the density profile at times T=0,50,100, and 150. Curves
are offset from one another for clarity. The heavy line shows a $\gamma=1$ slope.}
\end{figure}

\begin{figure}
\plotone{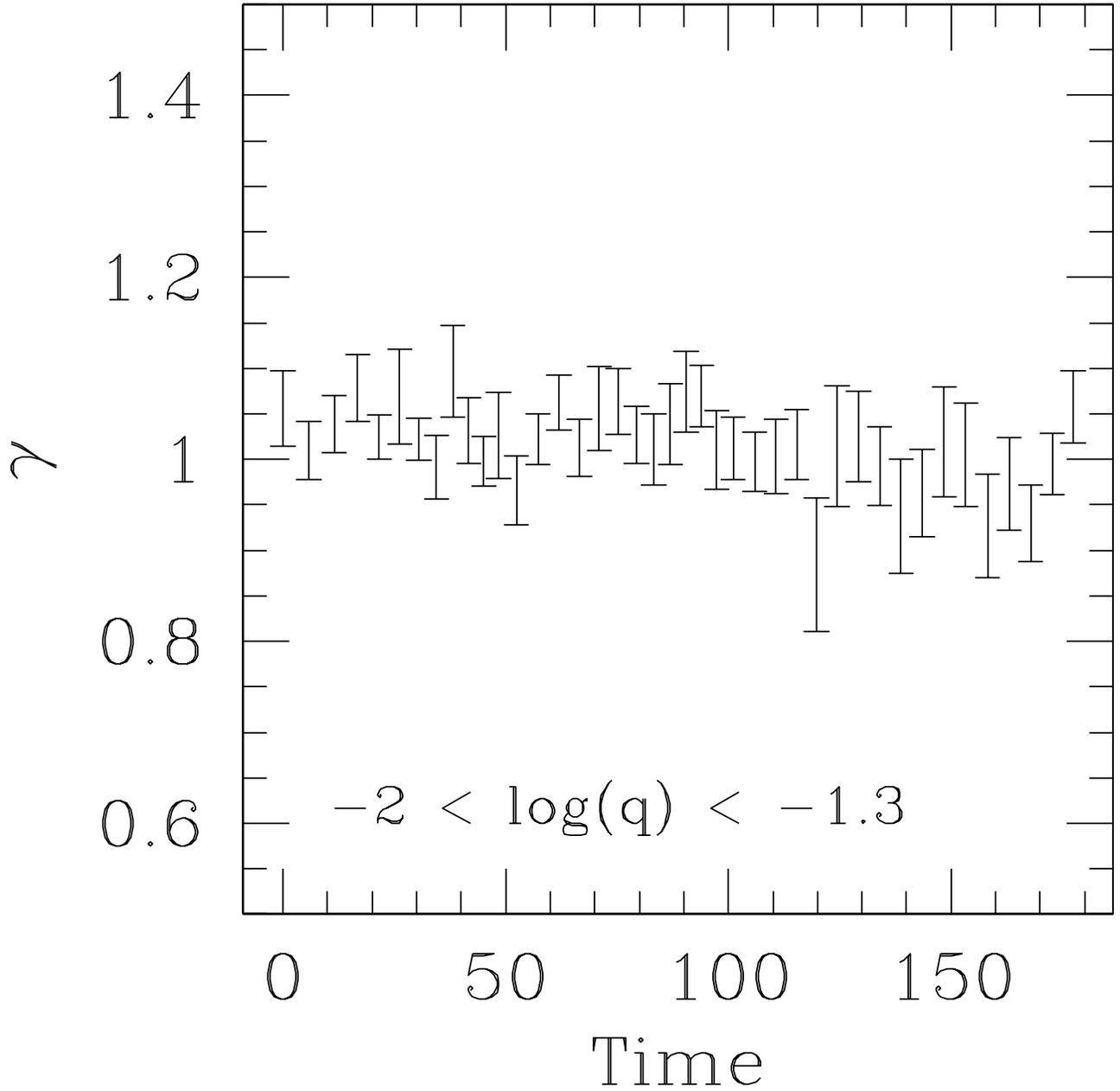}
\caption{The central density slope $\gamma$ of the model, fitted over
a range $-2 < \log(q) < -1.3$. The slope is slightly greater than
$\gamma=1$ at early times due to the fact that the density profile
steepens in the outer portion of the fitted range.}
\end{figure}

\begin{figure}
\plotone{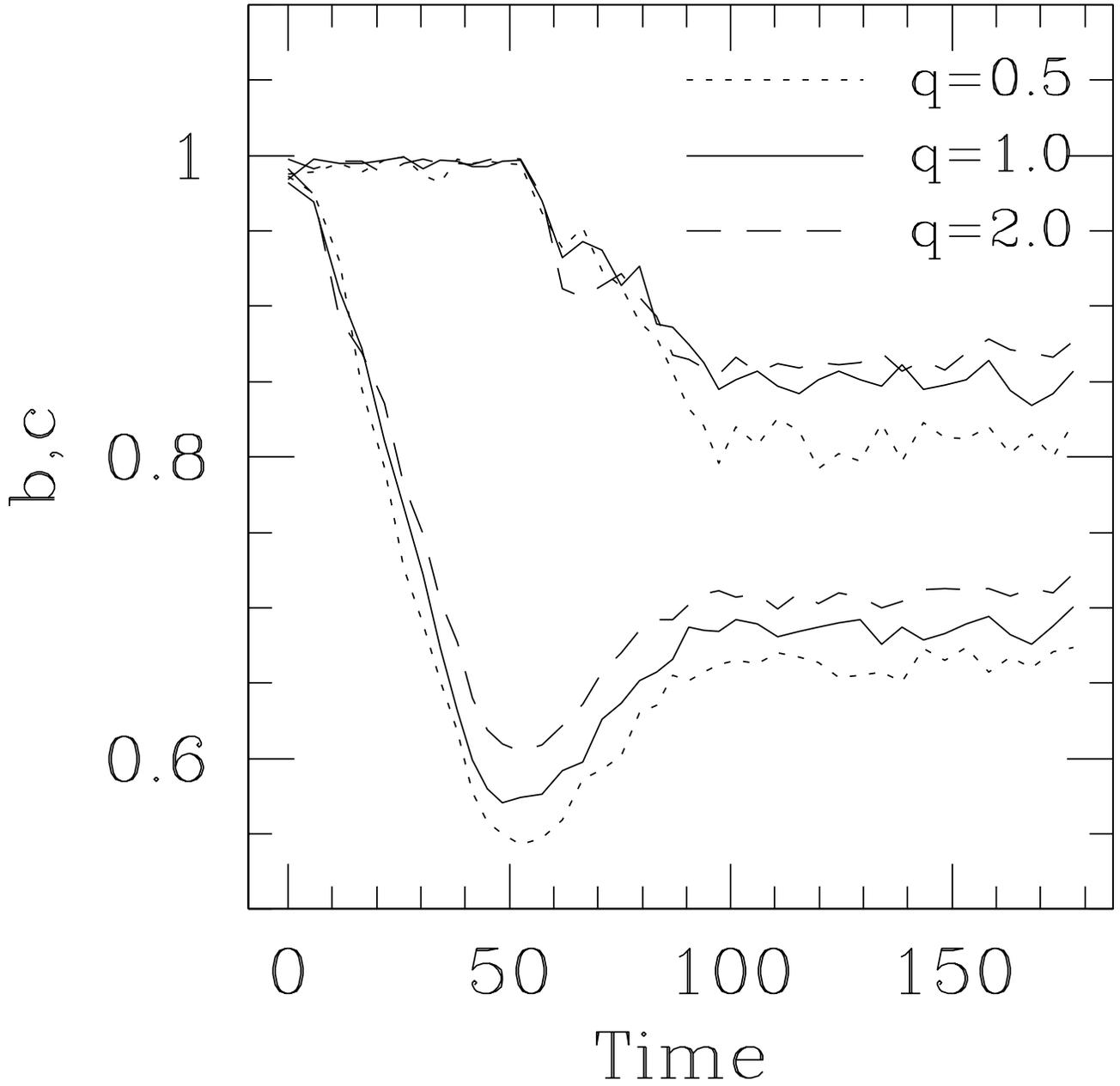}
\caption{The intermediate and minor axes lengths as a function of
time. The axes lengths are iteratively calculated from the ellipsoidal
density distribution using the moment of inertia tensor. See text for details.}
\end{figure}

\begin{figure}
\plotone{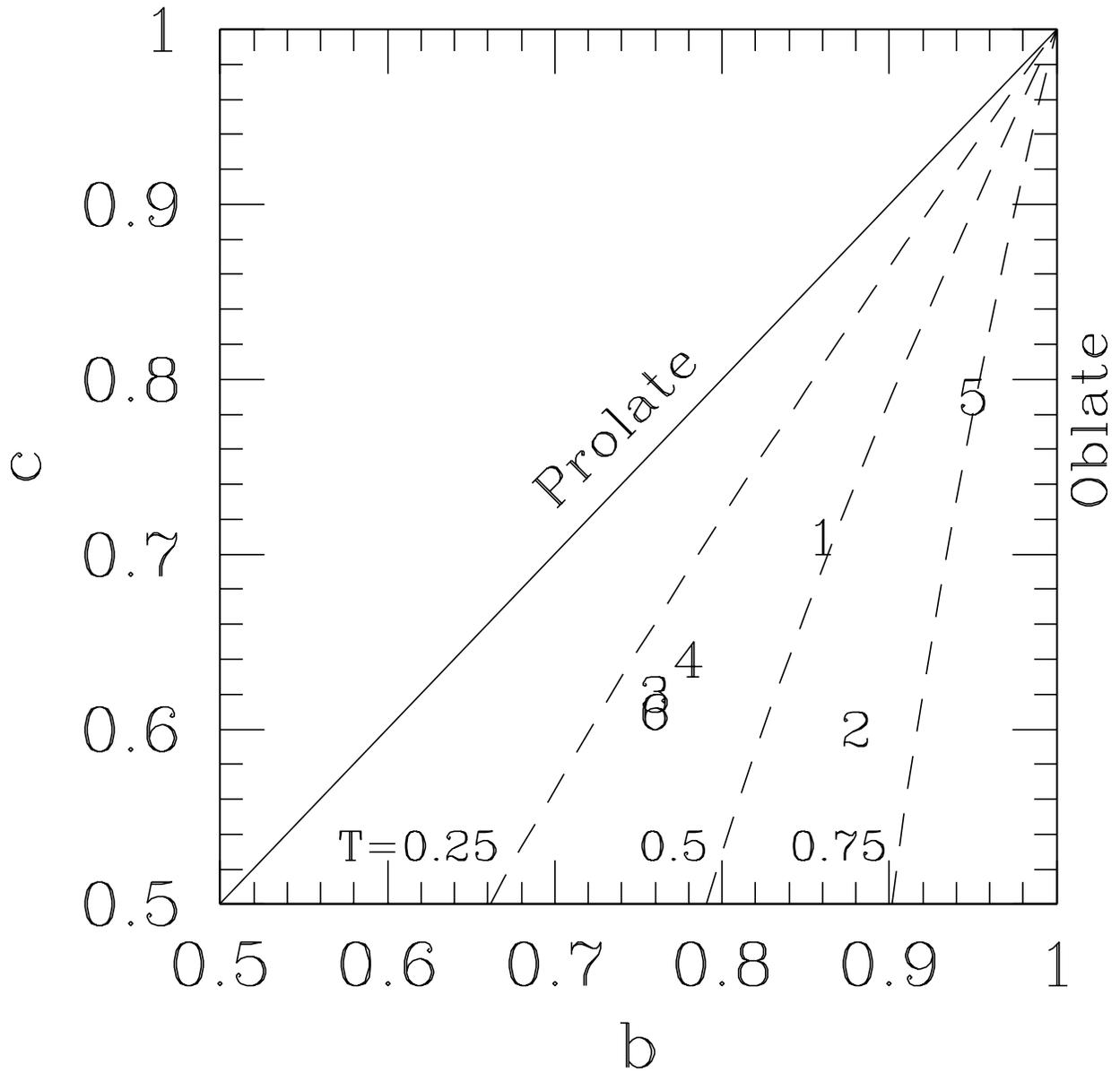}
\caption{The half-mass axes ratios of models generated with different
drag coefficients, plotted on the triaxiality plot of Franx \etal
(1991).}
\end{figure}

\begin{figure}
\plotone{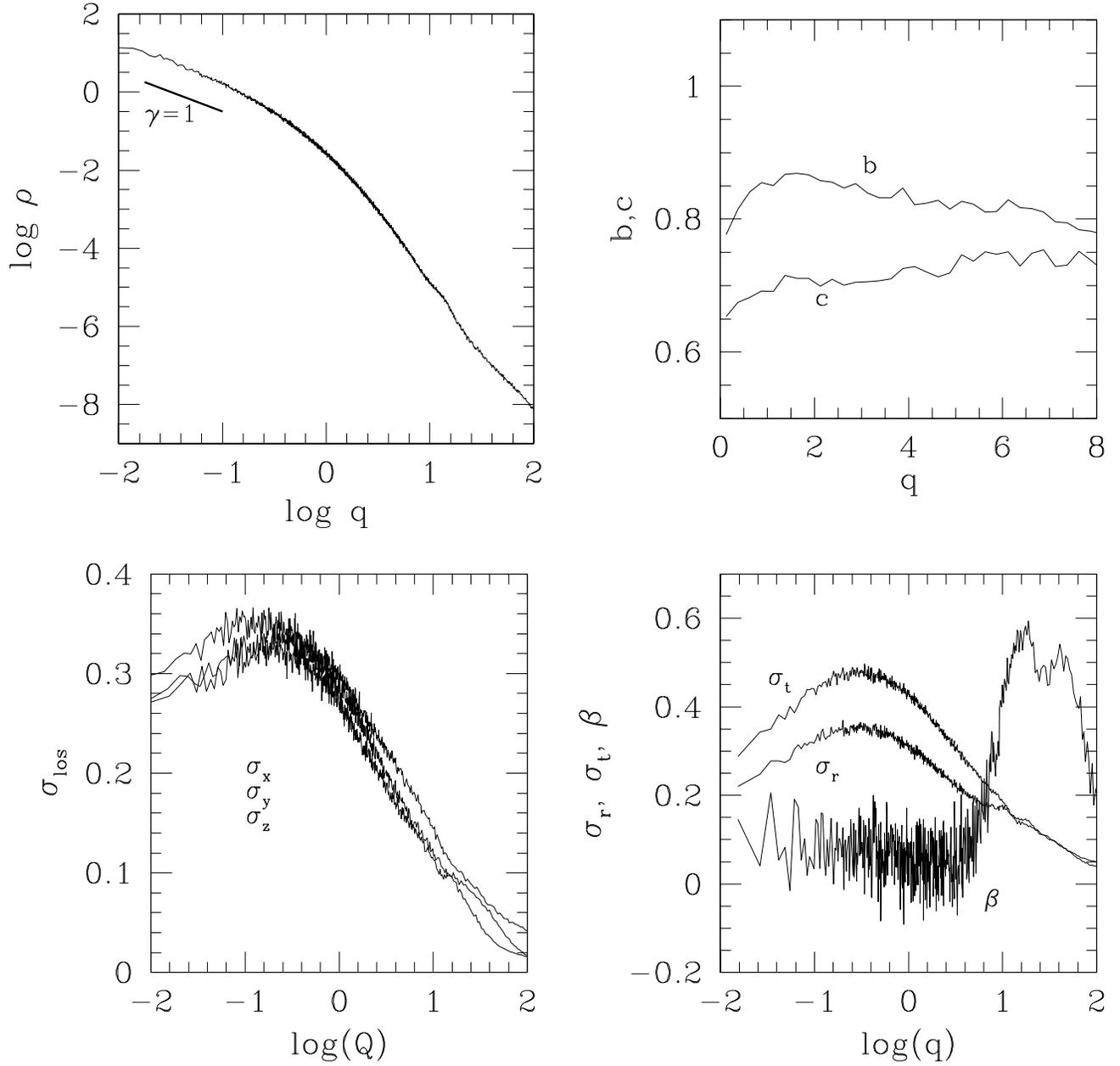}
\caption{The structural and kinematic properties of the model at $T=150$.
Upper left: density profile. Upper right:intermediate and minor axes lengths 
as a function of ellipsoidal radius. Lower left: projected velocity dispersion
along the fundamental axes, as a function of projected ellipsoidal radius.
Lower right: true radial and tangential velocity dispersion, and velocity
anisotropy parameter, as a function of ellipsoidal radius.}
\end{figure}

\begin{figure}
\plotone{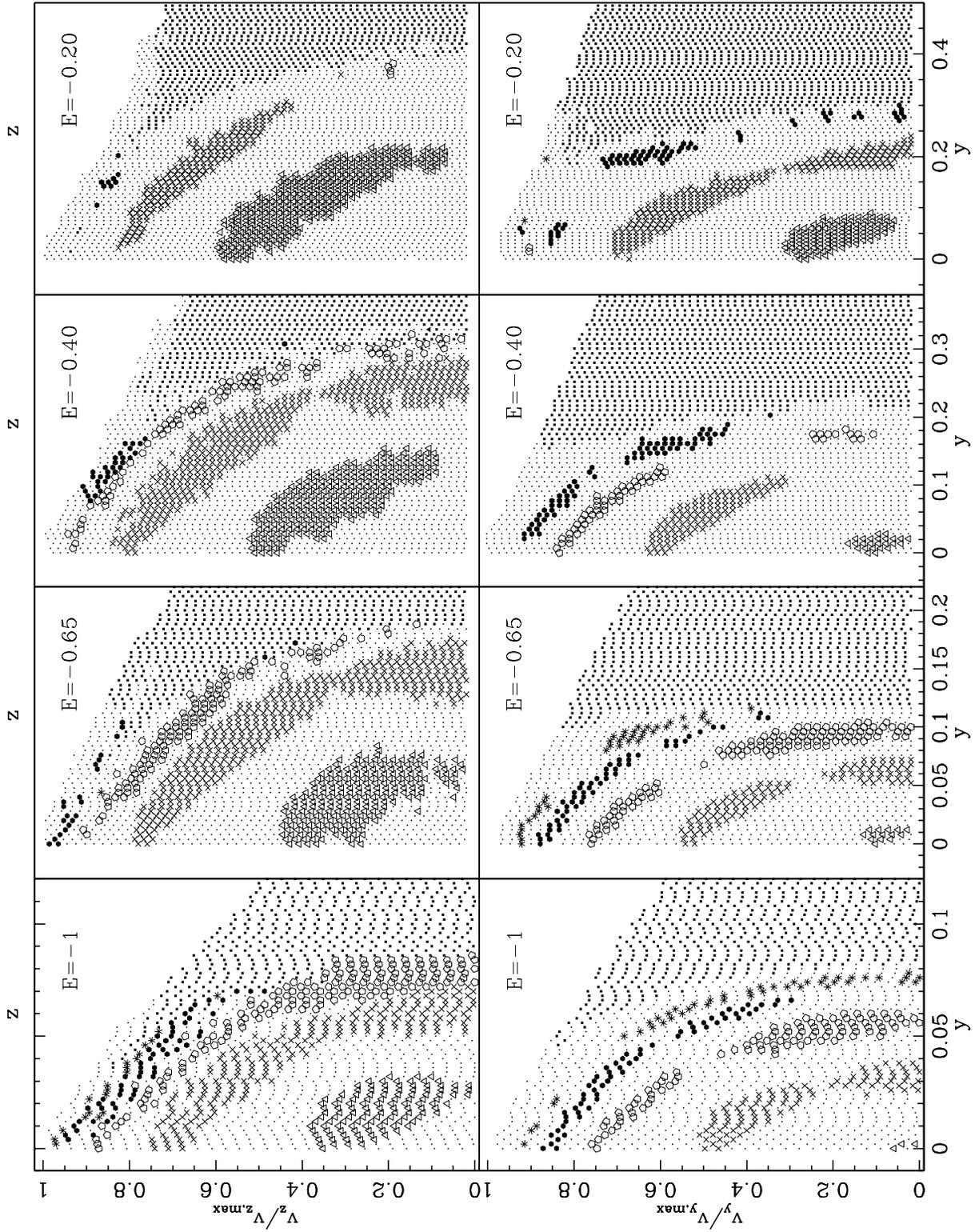}
\caption{Surfaces of section for the triaxial model at $T=150$,
plotted for orbital populations of differing binding energies. Top:
Surfaces of section for orbits in the $xz$ plane. Bottom: Surfaces of
section for orbits in the $xy$ plane. Orbits are coded by point type -
loops: filled boxes, boxes: small points, fish: open triangles, pretzels: X marks, 5:4 resonance: open pentagons, 6:5 resonance: filled hexagons, 
7:6 resonance: asterisks. This plot was created
by taking an average of all orbit types at a particular position on the 
surface of section. We zoom in on the x-axis of the plot to show as
many box and boxlets as possible; the orbits outside the limits of the plot 
are all tubes. If plotted to the full extent of the x axis, the
boxlet region would comprise $\sim 50\%$ of the most bound panels, and only $\sim 10\%$
of the least bound panels.}
\end{figure}

\begin{figure}
\plotone{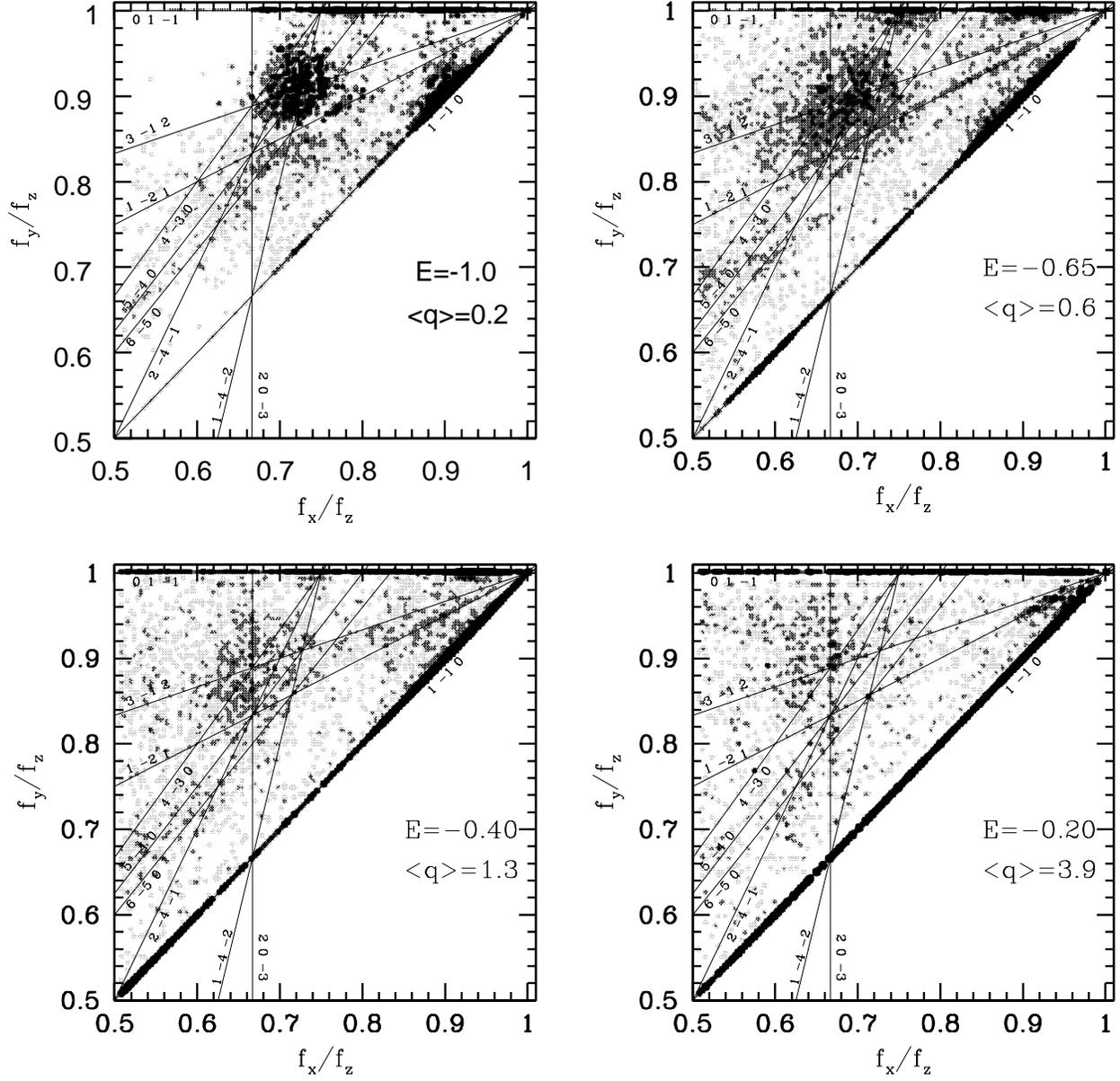}
\caption{Frequency map for the triaxial model at $T=150$,
plotted for orbital populations of differing binding energies. The
greyscale represents the number of orbits at a given frequency ratio. The lightest grey is 1 orbit, while black
is greater than 50 orbits.}
\end{figure}

\clearpage

\begin{deluxetable}{ccccc}
\tablecolumns{5}
\tablewidth{0pc}
\tablecaption{Triaxial Models}
\tablehead{
  \colhead{Model} & \colhead{$\xi_{0,z}$} & \colhead{$\xi_{0,y}$} & 
  \colhead{$b_{1\over 2}$} & \colhead{$c_{1\over 2}$} 
}
\startdata
1 & 0.030 & 0.025 & 0.85 & 0.70\\ 
2 & 0.030 & 0.045 & 0.87 & 0.59\\ 
3 & 0.045 & 0.045 & 0.75 & 0.61\\ 
4 & 0.045 & 0.030 & 0.77 & 0.63\\ 
5 & 0.010 & 0.020 & 0.94 & 0.78\\ 
6 & 0.045 & 0.020 & 0.75 & 0.60\\ 
\enddata
\end{deluxetable}

\end{document}